\documentstyle[12pt]{article}
\date{\small}
\title{\bf Accelerating Universe and Dynamical Compactification of Extra Dimensions}
\author{F. Darabi\thanks{e-mail:
f.darabi@azaruniv.edu}\\{\small Department of Physics, Azarbaijan
University of Tarbiat Moallem, 53714-161, Tabriz, Iran .}\\
{\small Institute for Theoretical and Mathematical Physics,
Niavaran, 19395-5746, Tehran, Iran.}}
\begin{document}
\maketitle
\begin{abstract}
We study a $(4+D)$-dimensional Kaluza-Klein cosmology with a
Robertson-Walker type metric having two scale factors $a$ and $R$,
corresponding to $D$-dimensional internal space and 4-dimensional
universe, respectively. By introducing an exotic matter as the
space-time part of the higher dimensional energy-momentum tensor,
a 4-dimensional decaying cosmological term is appeared as $\Lambda
\sim R^{-2}$, playing the role of an evolving dark energy in the
universe. The resulting field equations yield the exponential
solutions for the scale factors. These exponential behaviors may
account for the dynamical compactification of extra dimensions and
the accelerating expansion of the 4-dimensional universe in terms
of the Hubble parameter. The acceleration of universe may be
explained by the negative pressure of the exotic matter. It is
shown that the rate of compactification of higher dimensions
depends on the dimension, $D$. We then obtain the Wheeler-DeWitt
equation and find the general exact solutions in $D$-dimensions. A
good correspondence between the classical solutions and the
Wheeler-DeWitt solutions, in any dimension $D$, is obtained.
\end{abstract}
PACS number(s): 04.50.+h, 98.80.-k, 98.80.Qc
\newpage
\section{Introduction}
The study of $\Lambda$-decaying cosmological models has recently
been the subject of particular interest both from classical and
quantum aspects. A dynamical characteristic for the cosmological
term was already attributed by quantum field theorists since the
developments in particle physics and inflationary scenarios.
According to modern quantum field theory, the structure of a
vacuum turns out to be interrelated with some spontaneous
symmetry-breaking effects through the condensation of quantum
(scalar) fields. This phenomenon gives rise to a non-vanishing
vacuum energy density of the scale $M_p^4$ ($M_p$ is the plank
mass). Therefore, the bare cosmological constant may receive
potential contributions from this mass scale resulting in a large
effective cosmological constant which is too far from its present
upper band. This is known as the cosmological constant problem.
The $\Lambda$ decaying models may serve as potential candidates to
solve this problem by decaying the large value of the cosmological
constant $\Lambda$ to its present observed value.

Also, there are strong (astronomical) observational motivations
for considering cosmological models in which $\Lambda$ is
dynamically decreasing as $\Lambda\sim R^{-m}$. Some models assume
{\em a priori} a fixed value for the parameter $m$. The case
$m=2$, corresponding to the cosmic string matter, has mostly been
taken based on dimensional considerations by some authors
\cite{1}. The case $m \approx 4$ which resembles the ordinary
radiation has also been considered by some other authors \cite{2}.
A third group of authors have also studied the case $m=3$
corresponding to the ordinary matter \cite{3}. There are also some
other models in which the value of $m$ is not fixed {\em a priori}
and the numerical bounds on the value of $m$ is estimated by
observational data or obtained by calculation of the quantum
tunnelling rate \cite{4}. Other aspects of $\Lambda$-decaying
models have also been discussed with no specific numerical bounds
on $m$ \cite{5}.

On the other hand, the idea that our 4-dimensional universe might
have emerged from a higher dimensional space-time is now receiving
much attention \cite{6} where the compactification of higher
dimensions plays a key role. Meanwhile, the recent distance
measurements of type Ia supernova suggest strongly an accelerating
universe \cite{A}. This accelerating expansion is generally
believed to be driven by an energy source called {\it dark energy}
which provides negative pressure, such as a positive cosmological
constant \cite{q}, or a slowly evolving real scalar field called
{\it quintessence} \cite{!}. Moreover, the basic conclusion from
all previous observations that $\sim$ 70 percent of the energy
density of the universe is in a dark energy sector, has been
confirmed after the recent WMAP \cite{!!}.

To model a universe based on these considerations one may start
from a fundamental theory including both gravity and standard
model of particle physics. In this regards, it is interesting to
begin with ten or eleven-dimensional space-time of
superstring/M-theory, in which case one needs a compactification
of ten or eleven-dimensional supergravity theory where an
effective 4-dimensional cosmology undergoes acceleration. However,
it has been known for some time that it is difficult to derive
such a cosmology and has been considered that there is a no-go
theorem that excludes such a possibility, if one takes the
internal space to be time-independent and compact without boundary
\cite{!!!}. However, it has recently been shown that one may avoid
this no-go theorem by giving up the condition of time-independence
of the internal space; and a solution of the vacuum Einstein
equations with compact hyperbolic internal space has been proposed
based on this model\cite{&&}. Similar accelerating cosmologies can
also be obtained for SM2 and SD2 branes , not only for hyperbolic
but also for flat internal space\cite{&&&}.

On the other hand, from cosmological point of view, it is not so
difficult to find cosmological models in which the 4-dimensional
universe undergoes an accelerating expansion and the internal
space contracts with time, exhibiting the {\it dynamical
compactification} \cite{7}, \cite{8}, \cite{9}.

In \cite{9}, for instance, it is shown that using a more general
metric, as compared to ref \cite{7}, and introducing matter
without specifying its nature, the size of compact space evolves
as an inverse power of the radius of the universe. The
Friedmann-Robertson-Walker equations of the standard
four-dimensional cosmology is obtained using an effective pressure
expressed in terms of the components of the higher dimensional
energy-momentum tensor, and the negative value of this pressure
may explain the acceleration of our present universe.

To the author's knowledge the question of $\Lambda$-decaying
cosmological model has not received much attention in higher
dimensional Kaluza-Klein cosmologies. Moreover, the exotic matter
has not been considered as an alternative candidate to produce the
acceleration of the universe. The purpose of the present paper is
to study a $(4+D)$-dimensional Kaluza-Klein cosmology ,with an
extended Robertson-Walker type metric, in this context. As we are
concerned with cosmological solutions, which are intrinsically
time dependent, we may suppose that the internal space is also
time dependent. It is shown that by taking this higher dimensional
metric and introducing a 4-dimensional exotic matter, a decaying
cosmological term $\Lambda \sim R^{-2}$ is appeared as a type of
dark energy and the resulting field equations yield the
exponential solutions for the scale factors of the
four-dimensional universe and the internal space. These solutions
may account for the accelerating universe and dynamical
compactification of extra dimensions, driven by the negative
pressure of the exotic matter \footnote{A similar work \cite{@@}
has already been done in which the same extended FRW metric was
chosen with a radiation fluid occupying all the extended
space-time. They found an inflation for 3-dimensions and a
contraction for the $D$ remaining spatial dimensions. }. It should
be noted, however, that the solutions in principle describe
typical inflation rather than the recently observed acceleration
of the universe which is known to take place in an ordinary matter
dominated universe. Nevertheless, regarding the fact that about 70
percent of the total energy density of the universe is of dark
energy type with negative pressure, we may approximate the matter
content of the universe with almost dark energy and consider the
present model as a too simplified model of a real accelerating
universe.

The quantum cosmology of this model is also studied by obtaining
the Wheeler-DeWitt equation and finding its solutions. It is then
shown that a good correspondence exists between the classical and
quantum cosmological solutions.

The paper is organized as follows: In section {\bf 2}, we
introduce the model by taking a higher dimensional
Robertson-Walker type metric and a higher dimensional matter whose
non-zero part is a four-dimensional exotic matter. In section {\bf
3}, we obtain the Einstein equations for the two scale factors. In
section {\bf 4}, we solve the Einstein equations and obtain the
solutions. In section {\bf 5}, we derive the Wheeler-DeWitt
equation, and in section {\bf 6}, the exact solutions of the
Wheeler-DeWitt equation is obtained. Finally, in section {\bf 7},
we show a good correspondence between the classical and quantum
cosmology. The paper is ended with a conclusion.

\section{The model}

We start with the metric considered in \cite{10}, in which the
space-time is assumed to be of Robertson-Walker type having an
internal space with dimension $D$. We adopt a real chart $\{t,
r^{i}, \rho^{a}\}$ with $t$, $r^{i}$, and $\rho^{a}$ denoting the
time, space coordinates and internal space dimensions,
respectively. We, therefore, take\footnote{There is a little
difference between this metric and that of \cite{10}, in that here
the lapse function is generally considered as $N(t)$ instead of
taking $N=1$.}
\begin{equation}
ds^2=-N^2(t) dt^2+R^2(t)\frac{dr^i
dr^i}{(1+\frac{kr^2}{4})^2}+a^2(t) \frac{d\rho^a
d\rho^a}{(1+k^\prime \rho^2)}, \label{1}
\end{equation}
where $N(t)$ is the lapse function, $R(t)$ and $a(t)$ are the
scale factor of the universe and the radius of internal space,
respectively; $r^2 \equiv r^i r^i (i=1, 2, 3), \rho^2\equiv \rho^a
\rho^a (a=1, ... D)$, and $k, k^\prime=0, \pm1$. We assume the
internal space to be flat with compact topology $S^D$, which means
$k^\prime =0$. This assumption is motivated by the possibility of
the compact spaces to be flat or hyperbolic in ``{\it accelerating
cosmologies from compactification}'' scenarios, as discussed
above.

The form of energy-momentum tensor is dictated by Einstein's
equations and by the symmetries of the metric (\ref{1}).
Therefore, we may assume
\begin{equation}
T_{AB}=diag( -\rho, p, p, p, p_{_D}, p_{_D}, ..., p_{_D} ),
\label{3}
\end{equation}
where $A$ and $B$ run over both the space-time coordinates and the
internal space dimensions. Now, we examine the case for which the
pressure along all the extra dimensions vanishes, $p_{_D}=0$. In
so doing, we are motivated by the {\it brane world} scenarios
where the matter is to be confined to the 4-dimensional universe,
so that all components of $T_{AB}$ is set to zero but the
space-time components \cite{!&} and it means no matter escapes
through the extra dimensions.

We assume the energy-momentum tensor $T_{\mu \nu}$ of space-time
to be an exotic $\chi$ fluid with the equation of state
\begin{equation}
p_\chi =(\frac{m}{3}-1)\rho_\chi, \label{4}
\end{equation}
where $p_\chi$ and $\rho_\chi$ are the pressure and density of the
fluid, respectively and the parameter $m$ is restricted to the
range $0\leq m \leq 2$ \cite{12}. It is worth noting that the
equation of state (\ref{4}) with $0\leq m \leq 2$ resembles a
universe with negative pressure matter, violating the strong
energy condition \cite{15} and this violation is required for a
universe to be accelerated \cite{&&}.

Using standard techniques we obtain the scalar curvature
corresponding to the metric (\ref{1}) and then substitute the
result into the dimensionally extended Einstein-Hilbert action (
without higher dimensional cosmological term) plus a matter term
indicating the above mentioned exotic fluid. This leads to the
effective Lagrangian \footnote{We take the Planck units,
$G=c=\hbar=1$}
\begin{equation}
L=\frac{1}{2N}Ra^D{\dot{R}}^2+\frac{D(D-1)}{12N}R^3a^{D-2}{\dot{a}}^2+
\frac{D}{2N}R^2a^{D-1}\dot{R}\dot{a}-\frac{1}{2}kNRa^D+\frac{1}{6}N\rho_{\chi}
R^3a^D, \label{5}
\end{equation}
where a dot represents differentiation w.r.t $t$. We now take a
closed $(k=1)$ universe. It is easily shown that substituting the
equation of state (\ref{4}) into the continuity equation
\footnote{To obtain this equation we may use the contracted
Bianchi identity in (4+D) dimensions, namely $\nabla_M
G^{MN}=\nabla_M T^{MN}=0$ together with $T_{a b}=T_{\mu a}=0$
which gives rise to $\nabla_\mu T^{\mu \nu}=0.$ }
\begin{equation}
\dot{\rho}_\chi R+3(p_\chi+\rho_\chi)\dot{R}=0, \label{6}
\end{equation}
leads to the following behavior of the energy density in a closed
($k=1$) Friedmann-Robertson-Walker universe \cite{12}
\begin{equation}
\rho_\chi(R)=\rho_\chi(R_0)\left(\frac{R_0}{R}\right)^m_,
\label{7}
\end{equation}
where $R_0$ is the value of the scale factor at an arbitrary
reference time $t_0$.

Now, we  may define the cosmological term \cite{4}
\begin{equation}
\Lambda\equiv\rho_\chi(R), \label{8}
\end{equation}
which leads to
\begin{equation}
L=\frac{1}{2N}Ra^D{\dot{R}}^2+\frac{D(D-1)}{12N}R^3a^{D-2}{\dot{a}}^2+
\frac{D}{2N}R^2a^{D-1}\dot{R}\dot{a}-\frac{1}{2}NRa^D+\frac{1}{6}N\Lambda
R^3a^D, \label{9}
\end{equation}
where the cosmological term is now decaying with the scale
factor$R$ as
\begin{equation}
\Lambda(R)=\Lambda(R_0)\left(\frac{R_0}{R}\right)^m_. \label{10}
\end{equation}
Note that $\Lambda$ is now playing the role of an evolving dark
energy \cite{*} in 4-dimensions, because we did not consider
explicitly a $(4+D)$ dimensional cosmological term in the action,
and $\Lambda$ appears merely due to the specific choice of the
equation of state (\ref{4}) for the exotic matter. The decaying
$\Lambda$ term may also explain the smallness of the present value
of the cosmological constant since as the universe evolves from
its small to large sizes the large initial value of $\Lambda$
decays to small values.

Of particular interest, to the present paper, among the different
values of $m$ is $m=2$ which has some interesting implications in
reconciling observations with inflationary models \cite{13}, and
is consistent with quantum tunnelling \cite{4}.

\section{Einstein equations}

We take $m=2$ and set the initial values of $R_0$ and
$\Lambda(R_0)$
\begin{equation}
\Lambda(R_0)R^2_0=3 \:\:\:, \:\:\:\Lambda(R)=\frac{3}{R^2},
\label{11}
\end{equation}
leading to a positive cosmological
term which, according to (\ref{8}), guarantees the weak energy
condition $\rho_\chi >0$.

The lapse function $N(t)$, in principle, is also an arbitrary
function of time due to the reparametrization invariance of the
action. We, therefore, take the gauge
\begin{equation}
N(t)=R^3(t) a^D(t). \label{12}
\end{equation}
Now, the Lagrangian becomes
\begin{equation}
L=\frac{1}{2}\frac{\dot{R}^2}{R^2}+\frac{D(D-1)}{12}\frac{\dot{a}^2}{a^2}+\frac{D}{2}\frac{\dot{R}\dot{a}}{Ra},
\label{13}
\end{equation}
where Eq.(\ref{11}) has been used. It is seen that the parameters
$k$  and $\Lambda$ are effectively removed from the Lagrangian and
this implies that although $k$ and $\Lambda$ are not zero in this
model the corresponding 4-dimensional universe is equivalent to a
flat universe with a zero cosmological term. In other words, we do
not distinguish between our familiar 4-dimensional universe, which
seems to be flat and without any exotic fluid, and a closed universe
filled with an exotic fluid.\\
We now define the new variables
\begin{equation}
X=\ln{R}\:\:\:,\:\:\:Y=\ln{a}. \label{14}
\end{equation}
The lagrangian (\ref{13}) is written as
\begin{equation}
L=\frac{1}{2}\dot{X}^2+\frac{D(D-1)}{12}\dot{Y}^2+\frac{D}{2}\dot{X}\dot{Y}.
\label{15}
\end{equation}
The equations of motion are obtained
\begin{equation}
\ddot{X}+\frac{D}{2}\ddot{Y}=0, \label{16}
\end{equation}
\begin{equation}
\ddot{X}+\frac{D-1}{3}\ddot{Y}=0. \label{17}
\end{equation}
Combining the equations (\ref{16}) and (\ref{17}) we obtain
\begin{equation}
\ddot{X}=0, \label{18}
\end{equation}
\begin{equation}
\ddot{Y}=0. \label{19}
\end{equation}

\section{Solutions}

The solutions for $R(t)$ and $a(t)$ are as follows
\begin{equation}
R(t)=Ae^{\alpha t}, \label{20}
\end{equation}
\begin{equation}
a(t)=Be^{\beta t}, \label{21}
\end{equation}
where the constants $A$, $B$, $\alpha$ and $\beta$ should be
obtained, in principle, in terms of the initial conditions. It is
a reasonable assumption that the size of all spatial dimensions be
the same at $t=0$. Moreover, it may be assumed that this size
would be the Planck size $l_p$ in accordance with quantum
cosmological considerations. Therefore, we take $R(0)=a(0)={\it
l_P}$ so that $A=B={\it l_p}$, and
\begin{equation}
R(t)={\it l_p}e^{\alpha t}, \label{22}
\end{equation}
\begin{equation}
a(t)={\ l_p}e^{\beta t}. \label{23}
\end{equation}
It is important to note that the constants $\alpha, \beta$  are
not independent, and a relation may be obtained between them. This
is done by imposing the zero energy condition $H=0$ which is the
well-known result in cosmology as a {\it time reparametrization
invariant} theory. The Hamiltonian constraint is obtained through
the Legender transformation of the Lagrangian (\ref{15})
\begin{equation}
H=\frac{1}{2}\dot{X}^2+\frac{D(D-1)}{12}\dot{Y}^2+\frac{D}{2}\dot{X}\dot{Y}=0,
\label{24}
\end{equation}
which is written in terms of  $\alpha$ and $\beta$ as
\begin{equation}
H=\frac{1}{2}\alpha^2+\frac{D(D-1)}{12}\beta^2+\frac{D}{2}\alpha\beta=0.
\label{25}
\end{equation}
This constraint is satisfied only for $\alpha\le 0, \beta\ge 0$ or
$\alpha\ge 0, \beta\le 0$.

For $D \neq 1$, the case $\alpha=0$ or $\beta=0$ gives rise to
time independent scale factors, namely $R=a=l_P$, which is not
physically viable. We, therefore, choose $\alpha>0 , \beta<0$ so
that the universe and the internal space would expand and
contract, respectively, in accordance with the present
observations.

For the case $D=1$, we find
\begin{equation}
\left \{ \begin{array}{ll}
\beta=\mbox{arbitrary} \\
\alpha=0
\end{array}\right.
\:\: \mbox{or} \:\:\:\: \alpha=-\beta. \label{26}
\end{equation}
The former is not physically viable, since it predicts no time
evolution for the universe. The latter, however, may predict
exponential expansion for $R(t)$, and exponential contraction for
$a(t)$, both with the same exponent $\alpha>0$.

For the general case $D>1$, we find
\begin{equation}
\alpha_\pm=\frac{D\beta}{2}\left[-1\pm
\sqrt{1-\frac{2}{3}(1-\frac{1}{D})}\right], \label{27}
\end{equation}
which gives two positive values for $\alpha$ indicating two
possible expanding universes provided $\beta<0$ which indicates
the compactification of extra dimensions. Moreover, the values of
$\alpha_\pm$, for a given negative value of $\beta$, become larger
for higher dimensions. Therefore, the universe expands more
rapidly in both possibilities. On the contrary, for a given
positive value of $\alpha$, indicating an expanding universe, the
parameter $\beta$ may take two negative values
\begin{equation}
\beta_{\pm}=\frac{2 \alpha}{D}\left[-1\pm
\sqrt{1-\frac{2}{3}(1-\frac{1}{D})}\right]^{-1},
\end{equation}
indicating two ways of compactification. Moreover, they become
smaller for higher dimensions, exhibiting lower rates of
compactification.

To find the constants $\alpha, \beta$ we first obtain the Hubble
parameter for $R(t)$
\begin{equation}
H=\frac{\dot{R}}{R}=\alpha, \label{30}
\end{equation}
by which the constant $\alpha$ is fixed. The observed positive
value of $H$ will then justify our previous assumption,
$\alpha>0$. We may, therefore, write the solutions (\ref{22}) and
(\ref{23}) in terms of the Hubble parameter $H$ as
\begin{equation}
R(t)=l_p e^{Ht}, \label{30'}
\end{equation}
\begin{equation}
a(t)=l_p e^{-Ht}, \label{30''}
\end{equation}
for $D=1$, and
\begin{equation}
R(t)=l_p e^{Ht}, \label{31}
\end{equation}
\begin{equation}
a(t)_\pm=l_p e^{\frac{2Ht}{D}\left[-1\pm
\sqrt{1-\frac{2}{3}(1-\frac{1}{D})}\right]^{-1}}, \label{32}
\end{equation}
and
\begin{equation}
R_{\pm}(t)=l_p e^{\frac{D \beta t}{2}\left[-1\pm
\sqrt{1-\frac{2}{3}(1-\frac{1}{D})}\right]}, \label{0}
\end{equation}
\begin{equation}
a(t)=l_p e^{\beta t}. \label{00}
\end{equation}
for $D>1$.

For a given $H>0$, it is seen that the solution corresponding to
$D=1$ may predict an accelerating (de Sitter) universe and a
contracting internal space with exactly the same rates. For $D>1$,
in Eqs.(\ref{31}) and (\ref{32}), for a given $H>0$ in the
exponent of $R(t)$ the exponent in $a(t)$ takes two negative
values and becomes smaller for higher dimensions. This means that
while the 4-dimensional (de Sitter) universe is expanding by the
rate $H$, the higher dimensions may be compactified in two
possible ways with different rates of compactification as a
function of dimension, $D$. In Eqs.(\ref{0}) and (\ref{00}), on
the other hand, for a given $\beta<0$ the exponent in $R(t)$ takes
two positive values which become larger for higher dimensions.
This also means that while the extra dimensions contract by the
rate $\beta$, the universe may be expanded in two possible ways
with different expansion rates as a function of $D$.

It is easy to show that the Lagrangian (\ref{15}) ( or the
equations of motion ) is invariant under the simultaneous
transformation
\begin{equation}
R\rightarrow R^{-1}\:\:\:\:,\:\:\:\: a\rightarrow a^{-1},
\label{28}
\end{equation}
which is consistent with the time reversal $t\rightarrow -t$.
Therefore, four different phases of ``{\it
expansion-contraction}'' for $R(t)$ and $a(t)$ are distinguished,
Eqs.(\ref{31}) - (\ref{00}). One may prefer the ``{\it expanding
$R(t)$ - contracting $a(t)$}'' phase to ``{\it expanding $a(t)$ -
contracting $R(t)$}'' one, considering the present status of the
4D universe \footnote{For the special case $D=3$, both the
Lagrangian (\ref{15}) and the Hamiltonian constraint (\ref{24})
are invariant under the transformation
$$
a\rightarrow R \:\:\:\:,\:\:\:\: R \rightarrow a.
$$
Therefore, we have a dynamical symmetry between $R$ and $a$,
namely
$$
a \leftrightarrow R.
$$
In this case there is no real line of demarcation between $a$ and
$R$ to single out one of them as the real scale factor of the
universe. This is because the internal space is flat $k'=0$ and
according to (\ref{13}) one may assume the 4D universe with $k,
\Lambda\neq0$ to be equivalent to the one in which $k=\Lambda=0$.
Therefore, both have the same topology $S^3$.}.

The deceleration parameter $q$ for the scale factor $R$ is
obtained
\begin{equation}
q=-\frac{\ddot{R}R}{\dot{R}^2}=-1. \label{29}
\end{equation}
Observational evidences not only do not rule out the negative
deceleration parameter but also puts the limits on the present
value of $q$ as $-1\leq q < 0$ \cite{A}. Therefore, this negative
value seems to favor a cosmic acceleration in the expansion of the
universe.

In the expansion phase of the closed ($k=1$) universe the
cosmological term $\Lambda$ decays exponentially with time $t$ as
\begin{equation}
\Lambda(t)=3l_p^{-2}e^{-2Ht}, \label{33}
\end{equation}
whereas in the contraction phase ($t \rightarrow -t$) it grows
exponentially to large values so that at $t=0$ it becomes
extremely large, of the order of $M_p^2$. This huge value of
$\Lambda$ may be extinguished rapidly by assuming a sufficiently
large Hubble parameter $H$, consistent with the present
observations, to alleviate the cosmological constant problem.

\section{Wheeler-DeWitt equation}

To obtain the Wheeler-DeWitt equation we start with the Lagrangian
(\ref{15}). The conjugate momenta corresponding to $X$ and $Y$ are
obtained
\begin{equation}
P_X=\frac{\partial L}{\partial
\dot{X}}=\dot{X}+\frac{D}{2} \dot{Y},
\end{equation}
\begin{equation}
P_Y=\frac{\partial L}{\partial
\dot{Y}}=\frac{D}{2}\dot{X}+\frac{D(D-1)}{6}\dot{Y}, \label{34}
\end{equation}
from which we obtain
\begin{equation}
\dot{X}=\frac{6}{D+2}\left[P_X
\left(\frac{1-D}{3}\right)+P_Y\right], \label{34'}
\end{equation}
\begin{equation}
\dot{Y}=\frac{6}{D(D-1)}\left[P_Y\frac{2(1-D)}{D+2}-P_X\frac{D(1-D)}{D+2}\right].
\label{35}
\end{equation}
Substituting Eqs.(\ref{34'}), (\ref{35}) into the Hamiltonian
constraint (\ref{24}), we obtain
\begin{equation}
H=(1-D)P_X^2-\frac{6}{D}P_Y^2+6P_X P_Y=0. \label{36}
\end{equation}
Now, we may use the following quantum mechanical replacements
$$
P_X \rightarrow -i\frac{\partial}{\partial X}\:\:\:\:,\:\:\:\: P_Y
\rightarrow -i\frac{\partial}{\partial Y},
$$
by which the Wheeler-DeWitt equation is obtained
\begin{equation}
\left[(D-1)\frac{\partial^2}{\partial
X^2}+\frac{6}{D}\frac{\partial^2}{\partial
Y^2}-6\frac{\partial}{\partial X}\frac{\partial}{\partial
Y}\right]\Psi(X, Y)=0, \label{37}
\end{equation}
where $\Psi(X, Y)$ is the wave function of the universe in the
$(X, Y)$ mini-superspace.

We introduce the following change of variables
\begin{equation}
x=X(1-\frac{D}{D+3})+\frac{D}{D+3}Y \:\:\:\:,\:\:\:\:
y=\frac{X-Y}{D+3}, \label{38}
\end{equation}
by which the Wheeler-DeWitt equation takes a simple form
\begin{equation}
\left\{ -3\frac{\partial^2}{\partial x^2}+\frac{D+2}{D}
\frac{\partial^2}{\partial y^2}\right\} \Psi(x, y)=0. \label{39}
\end{equation}
Now, we can separate the variables as $\Psi(x, y)=\phi(x) \psi(y)$
to obtain the following equations
\begin{equation}
\frac{\partial^2 \phi(x)}{\partial x^2}=\frac{\gamma}{3} \phi(x),
\label{40}
\end{equation}
\begin{equation}
\frac{\partial^2 \psi(y)}{\partial y^2}=\frac{\gamma
D}{D+2}\psi(y), \label{41}
\end{equation}
where we assume $\gamma >0$.

\section{Solutions}

The solutions of Eqs.(\ref{40}), (\ref{41}) in terms of $x, y$ are
as follows
\begin{equation}
\phi(x)=e^{\pm\sqrt{\frac{\gamma}{3}}x}, \label{42}
\end{equation}
\begin{equation}
\psi(y)=e^{\pm\sqrt{\frac{\gamma D}{D+2}}y}, \label{43}
\end{equation}
leading to the four possible solutions for $\Psi(x, y)$ as
\begin{equation}
\Psi_D^{\pm} (x, y)= A^{\pm}e^{\pm\sqrt{\frac{\gamma}{3}}x \pm
\sqrt{\frac{\gamma D}{D+2}}y}, \label{43'}
\end{equation}
\begin{equation}
\Psi_D^{\pm} (x, y)= B^{\pm}e^{\pm\sqrt{\frac{\gamma}{3}}x \mp
\sqrt{\frac{\gamma D}{D+2}}y}, \label{44}
\end{equation}
or alternative solutions in terms of $X, Y$ as
\begin{equation}
\Psi_D^{\pm} (X, Y)= A^{\pm}e^{\pm \sqrt{\frac{\gamma}{3}} \left(
\frac{3X+DY}{D+3}\right) \pm \sqrt{\frac{\gamma D}{D+2}}\left(
\frac{X-Y}{D+3}\right)}, \label{45}
\end{equation}
\begin{equation}
\Psi_D^{\pm} (X, Y)= B^{\pm}e^{\pm \sqrt{\frac{\gamma}{3}} \left(
\frac{3X+DY}{D+3}\right) \mp \sqrt{\frac{\gamma
D}{D+2}}\left(\frac{X-Y}{D+3}\right)}, \label{46}
\end{equation}
where $A^{\pm}, B^{\pm}$ are the normalization constants. We may
also write down the solutions in terms of $R$ and $a$
\footnote{For $D=3$, there is a exchange symmetry $\Psi(R, a)
\leftrightarrow \Psi(a, R)$ under the exchange $ a \leftrightarrow
R$.}
\begin{equation}
\Psi_D^{\pm} (R, a)= A^{\pm} R^{\pm
\frac{1}{D+3}\left(\sqrt{3\gamma}+\sqrt{\frac{\gamma
D}{D+2}}\right)} a^{\pm \frac{1}{D+3} \left(
\sqrt{\frac{\gamma}{3}}D-\sqrt{\frac{\gamma D}{D+2}}\right)},
\label{47}
\end{equation}
\begin{equation}
\Psi_D^{\pm} (R, a)= B^{\pm} R^{\pm
\frac{1}{D+3}\left(\sqrt{3\gamma}-\sqrt{\frac{\gamma
D}{D+2}}\right)} a^{\pm \frac{1}{D+3}
\left(\sqrt{\frac{\gamma}{3}}D+\sqrt{\frac{\gamma
D}{D+2}}\right)}. \label{48}
\end{equation}

It is now important to impose the {\it good} boundary conditions
on the above solutions to single out the physical ones. In so
doing, we may impose the following condition
\begin{equation}
\Psi_D(R\rightarrow \infty, a\rightarrow \infty)=0.
\end{equation}
Then, one may take the following solutions
\begin{equation}
\Psi_D^{\pm} (R, a)= C^{\pm} R^{-
\frac{1}{D+3}\left(\sqrt{3\gamma}\pm\sqrt{\frac{\gamma
D}{D+2}}\right)} a^{-\frac{1}{D+3} \left(\sqrt{\frac{\gamma}{3}}D
\mp \sqrt{\frac{\gamma D}{D+2}}\right)}, \label{49}
\end{equation}
where $C^{\pm}$ are the normalization constants and the exponents
of $R$ and $a$ are negative for any value of $D$ \footnote{For
$D=1$, the exponent of ``$a$'' corresponding to $\Psi^+$ becomes
zero so that $\Psi^+$ depends only on $R$ with the condition
$\Psi^+(R \rightarrow \infty) \rightarrow 0$.}.

One may obtain the solutions (\ref{49}) in $(X, Y)$
mini-superspace as
\begin{equation}
\Psi_D^{\pm} (X, Y)= C^{\pm}e^{-\sqrt{\frac{\gamma}{3}} \left(
\frac{3X+DY}{D+3}\right) \mp \sqrt{\frac{\gamma D}{D+2}}\left(
\frac{X-Y}{D+3}\right)}. \label{50}
\end{equation}

\section{Classical limits}

One of the most interesting topics in the context of quantum
cosmology is the mechanisms through which  classical cosmology may
emerge from quantum cosmology. When does a Wheeler-DeWitt wave
function predict a classical space-time?  Hartle has put forward a
simple rule for applying quantum mechanics to a single system
(universe): {\it If the wave function is sufficiently peaked about
some region in the configuration space we predict to observe a
correlation between the observables which characterize this
region.}

In this regards, we first take $D=1$ and look for a correspondence
between classical and quantum solutions. Using Eqs.(\ref{30'}) and
(\ref{30''}) in the Planck units, the corresponding classical
locus in $(R, a)$ configuration space, is
\begin{equation}
Ra=1, \label{51}
\end{equation}
whereas in $(X, Y)$ coordinates we have
\begin{equation}
X+Y=0. \label{52}
\end{equation}
We now consider the wave functions (\ref{50}) in $(X, Y)$
mini-superspace for $D=1$
\begin{equation}
\Psi^+_1(X, Y)=C^+ e^{-\sqrt{\frac{\gamma}{3}}X}, \label{53}
\end{equation}
\begin{equation}
\Psi^-_1(X, Y)=C^- e^{-\sqrt{\frac{\gamma}{3}}\frac{X+Y}{2}}.
\label{54}
\end{equation}
The above wave functions, in their present form, are not square
integrable as is required for the wave functions to predict the
classical limit. However, one may take the absolute value of the
exponents to make the wave functions square integrable
\begin{equation}
\Psi^+_1(X, Y)=C^+ e^{-|\sqrt{\frac{\gamma}{3}}X|}, \label{55}
\end{equation}
\begin{equation}
\Psi^-_1(X, Y)=C^- e^{-|\sqrt{\frac{\gamma}{3}}\frac{X+Y}{2}|}.
\label{56}
\end{equation}
We next consider the general case $D>1$. Eliminating the parameter
$t$ in Eqs. (\ref{31}) and (\ref{32}) the classical loci in terms
of $R, a$ are obtained
\begin{equation}
a_\pm=R^{\frac{2}{D}\left[-1\pm
\sqrt{1-\frac{2}{3}(1-\frac{1}{D})}\right]^{-1} }. \label{57}
\end{equation}
The corresponding forms of these loci in terms of $X, Y$ are
\begin{equation}
Y_+=\frac{2}{D}X\left[-1+
\sqrt{1-\frac{2}{3}(1-\frac{1}{D})}\right]^{-1}, \label{58}
\end{equation}
\begin{equation}
Y_-=\frac{2}{D}X\left[-1-
\sqrt{1-\frac{2}{3}(1-\frac{1}{D})}\right]^{-1}. \label{59}
\end{equation}
The wave functions (\ref{50}) also are not square integrable, so
we may replace the exponents by their absolute values
\begin{equation}
\Psi_D^{\pm} (X, Y)= C^{\pm}e^{-\left|\sqrt{\frac{\gamma}{3}}
\left( \frac{3X+DY}{D+3}\right) \mp \sqrt{\frac{\gamma
D}{D+2}}\left( \frac{X-Y}{D+3}\right)\right|}, \label{60}
\end{equation}
to make them square integrable. Now, following Hartle's point of
view, we try to make correspondence between the classical loci and
the wave functions.

Figures 1, 2, and 3 show respectively the 2D plots of the typical
wave functions $\Psi^+_1$, $\Psi^+_3$ and $\Psi^+_9$ in terms of
$(X, Y)$ for $\gamma=10^{-6}$; Figures 13, 14, and 15 show the
corresponding 3D plots, respectively. On the other hand, Figures
7, 8, and 9 show the classical loci (\ref{52}) for $D=1$,
(\ref{58}) for $D=3$ and $D=9$, respectively. It is seen that the
2D and 3D plots of the wave functions $\Psi^+_1$, $\Psi^+_3$ and
$\Psi^+_9$ are exactly peaked on the classical loci.

In the same way, Figures 4, 5, and 6 show respectively the 2D
plots of the wave functions $\Psi^-_3$, $\Psi^-_6$ and $\Psi^-_9$.
Figures 16, 17, and 18 show the corresponding 3D plots,
respectively. Figures 10, 11, and 12 show the classical loci
(\ref{59}) for $D=3$, $D=6$ and $D=9$, respectively. It is seen
again an exact correspondence between the 2D and 3D plots of the
wave functions $\Psi^-_3$, $\Psi^-_6$, $\Psi^-_9$ and the
classical loci. This procedure applies for all $D$.

\newpage
\section*{Conclusion}

In this paper, we have studied a $(4+D)$-dimensional Kaluza-Klein
cosmology with two scale factors, $R$ for the universe and $a$ for
the higher dimensional space. By introducing a typical exotic
matter in 4-dimensions, resulting in a decaying cosmological term,
we obtain exponential solutions for $R$ and $a$ in terms of the
Hubble parameter, indicating the accelerating expansion of the
universe and dynamical compactification of extra dimensions,
respectively. The rate of compactification is shown to be
dependent on the number of extra dimensions, $D$. The more extra
dimensions, the less rate of compactification. It is worth noting
that the model is free of initial singularity problem because both
$R$ and $a$ are non-zero at $t=0$, resulting in a finite Ricci
scalar.

The model, although describes in principle a closed universe with
non-vanishing cosmological constant, but is equivalent to a flat
universe with zero cosmological constant. Therefore, one may
assume that we are really living in a closed universe with
$\Lambda\neq 0$ , but it effectively appears as a flat universe
with $\Lambda= 0$. Note that we have not considered ordinary
matter sources in the model except an exotic matter source
which is to be considered as a source of dark energy. Therefore,
it seems the solutions to describe
typical inflation rather than the recently observed acceleration
of the universe which is known to take place in an ordinary matter
dominated universe.
However,
if the large percent of the matter sources in the universe would
be of dark energy type (as the present observations strongly
recommend), then one may keep the results here even in the
presence of other matter source, keeping in mind that the relevant
contribution to the total matter source of the universe is the
dark energy.

A question may arise on the fact that no physics is supposed to
exist below the planck length whereas for the contracting
solution, the scale factor $a(t)$ goes to zero starting from
$l_p$. However, it is not a major problem because we have not
considered elements of quantum gravity theory in this model and
merely studied a model based on general relativity which is
supposed to be valid in any scale without limitation. The scale
$l_p$, in this paper, is not introduced within a quantum gravity
model (action); it just appears as a typical initial condition, in
the middle of a classical model, based on the quantum cosmological
consideration. One may choose another scale based on some other
physical considerations.

We have also studied the corresponding quantum cosmology and
obtained the exact solutions. We have shown a good correspondence
between the classical and quantum cosmological solutions for any
$D$, provided that the wave functions vanish for the infinite
scale factors. There is no such a correspondence if another
boundary condition, other than stated, is taken.

\section*{Acknowledgment}

The author would like to thank A. Rezaei-Aghdam and referees for useful
and constructive comments. This work  has been supported by the Research office of
Azarbaijan University of Tarbiat Moallem, Tabriz, Iran.

\newpage
{\large {\bf Figure captions}} \vspace{10mm}
\\
FIG. 1. 2D plot of $\Psi^+_1$ in terms of $(X, Y)$ for
$\gamma=10^{-6}$
\\
FIG. 2. 2D plot of $\Psi^+_3$ in terms of $(X, Y)$ for
$\gamma=10^{-6}$
\\
FIG. 3. 2D plot of $\Psi^+_9$ in terms of $(X, Y)$ for
$\gamma=10^{-6}$
\\
FIG. 4. 2D plot of $\Psi^-_3$ in terms of $(X, Y)$ for
$\gamma=10^{-6}$
\\
FIG. 5. 2D plot of $\Psi^-_6$ in terms of $(X, Y)$ for
$\gamma=10^{-6}$
\\
FIG. 6. 2D plot of $\Psi^-_9$ in terms of $(X, Y)$ for
$\gamma=10^{-6}$
\\
FIG. 7. Classical locus $X+Y=0$ for $D=1$
\\
FIG. 8. Classical locus $Y_+=\frac{2}{D}X[-1+
\sqrt{1-\frac{2}{3}(1-\frac{1}{D})}]^{-1}$ for $D=3$
\\
FIG. 9. Classical locus $Y_+=\frac{2}{D}X[-1+
\sqrt{1-\frac{2}{3}(1-\frac{1}{D})}]^{-1}$ for $D=9$
\\
FIG. 10. Classical locus $Y_-=\frac{2}{D}X[-1-
\sqrt{1-\frac{2}{3}(1-\frac{1}{D})}]^{-1}$ for $D=3$
\\
FIG. 11. Classical locus $Y_-=\frac{2}{D}X[-1-
\sqrt{1-\frac{2}{3}(1-\frac{1}{D})}]^{-1}$ for $D=6$
\\
FIG. 12. Classical locus $Y_-=\frac{2}{D}X[-1-
\sqrt{1-\frac{2}{3}(1-\frac{1}{D})}]^{-1}$ for $D=9$
\\
FIG. 13. 3D plot of $\Psi^+_1$ in terms of $(X, Y)$ for
$\gamma=10^{-6}$
\\
FIG. 14. 3D plot of $\Psi^+_3$ in terms of $(X, Y)$ for
$\gamma=10^{-6}$
\\
FIG. 15. 3D plot of $\Psi^+_9$ in terms of $(X, Y)$ for
$\gamma=10^{-6}$
\\
FIG. 16. 3D plot of $\Psi^-_3$ in terms of $(X, Y)$ for
$\gamma=10^{-6}$
\\
FIG. 17. 3D plot of $\Psi^-_6$ in terms of $(X, Y)$ for
$\gamma=10^{-6}$
\\
FIG. 18. 3D plot of $\Psi^-_9$ in terms of $(X, Y)$ for
$\gamma=10^{-6}$
\end{document}